\documentclass[aps,prb,twocolumn,superscriptaddress,amsmath,amssymb,showpacs]{revtex4}
\usepackage{amssymb}
\usepackage[dvips]{graphicx}
\usepackage{bm}
\usepackage{epsfig}
\usepackage[latin1]{inputenc}

\begin{document}

\title{Magnetic structure of Sm$_2$IrIn$_8$}

\author{C. Adriano}
\email{cadriano@ifi.unicamp.br}
\affiliation{Instituto de F\'isica "Gleb Wataghin",UNICAMP,13083-970, Campinas-São Paulo, Brazil.}

\author{R. Lora-Serrano}
\affiliation{Instituto de F\'isica "Gleb Wataghin",UNICAMP,13083-970, Campinas-São Paulo, Brazil.}

\author{C. Giles}
\affiliation{Instituto de F\'isica "Gleb Wataghin",UNICAMP,13083-970, Campinas-São Paulo, Brazil.}

\author{F. de Bergevin}
\affiliation{European Synchrotron Radiation Facility, Grenoble 38043, France.}

\author{J. C. Lang}
\affiliation{Advanced Photon Source, Argonne National Laboratory, Argonne, Illinois 60439.}

\author{G. Srajer}
\affiliation{Advanced Photon Source, Argonne National Laboratory, Argonne, Illinois 60439.}

\author{C. Mazzoli}
\affiliation{European Synchrotron Radiation Facility, Grenoble 38043, France.}

\author{L. Paolasini}
\affiliation{European Synchrotron Radiation Facility, Grenoble 38043, France.}

\author{P. G. Pagliuso}
\affiliation{Instituto de F\'isica "Gleb Wataghin",UNICAMP,13083-970, Campinas-São Paulo, Brazil.}

\date{\today}

\begin{abstract}

The magnetic structure of the intermetallic antiferromagnet Sm$_2$IrIn$_8$ was determined using x-ray resonant
magnetic scattering (XRMS). Below $T_N$ = 14.2, Sm$_2$IrIn$_8$ has a commensurate antiferromagnetic
structure with a propagation vector $\vec{\eta} = (1/2,0,0)$. The Sm magnetic moments lie in the
\textit{ab} plane and are rotated roughly 18º away from the \textit{a} axis. The magnetic structure of this compound was obtained by measuring the strong dipolar resonant peak whose enhancement was of over
two orders of magnitude at the $L_2$ edge. At the $L_3$ edge both quadrupolar and dipolar features were observed in the energy line shape. The magnetic structure and properties of
Sm$_2$IrIn$_8$ are found to be consistent with the general trend already seen for the Nd-, Tb- and the
Ce-based compounds from the R$_{m}$M$_n$In$_{3m+2n}$ family (R = rare earth; M=Rh or Ir, $m$ = 1, 2; \textit{n} = 0, 1), where the crystalline electrical field (CEF) effects determine the direction
of magnetic moments and the $T_N$ evolution in the series. The measured N\'eel temperature for Sm$_2$IrIn$_8$ is slightly suppressed when compared to the $T_N$ of the parent cubic compound SmIn$_3$. 

\end{abstract}


\maketitle


\section{\bf INTRODUCTION}

The microscopic details of $4f$-electron magnetism play a fundamental role in the physical properties of various classes of
rare-earth based materials such as heavy fermions, magnetically
ordered alloys and permanent magnets. The existence of structurally related families of rare-earth based
compounds provides a great opportunity to explore how the
details of the $4f$-electrons magnetism evolve as a function of
changes in the dimensionality, local symmetry and electronic
structure along each related family. The recently
discovered\cite{Hegger,Petrovic1,Petrovic2,ThompsonJMMM,pagliuso3,pagliuso4,raimundo2,japa218}
family of intermetallic compounds R$_{m}$M$_n$In$_{3m+2n}$ (M = Co,
Rh or Ir, $m$ = 1, 2; $R$ = La, Ce, Pr, Nd, Sm, Gd) have proved to be very promising in this regard,
since it possesses many members of structurally related
heavy-fermions superconductors (HFS), for R = Ce, antiferromagnets (R
= Nd, Sm, Gd and Tb) and paramagnetic metals (R = La, Pr). Within
this family, the physical properties of a particular R-member can
also be compared to compounds based on the same R with three different
related structures [the cubic RIn$_{3}$ and the tetragonal
RMIn$_{5}$(1-1-5) and R$_2$MIn$_8$
(2-1-8)]\cite{Moshopoulou1,Moshopoulou2,buschow} and/or to the same R formed with
three distinct transition metals (M = Rh, Ir and Co - not for all
R -) in the same structure.

For the Ce-based HFS in this family, extensive investigation has
revealed fascinating physical properties such as quantum
criticality, non-fermi-liquid-behavior and an intriguing interplay
between magnetism and superconductivity, reflected in very rich
phase
diagrams.\cite{pagliuso1,pagliuso2,Fisk3,Zapf,ThompsonNature,Sidorov,Bianchi,eric2,paglione} Because the HFS members of this family are structurally related,
its investigation has been used to provide some insights on
the question why some structure types are favorable to host many
superconductors. A possible relationship between the
superconducting critical temperature T$_{c}$ and the crystalline
anisotropy\cite{pagliuso2,Kumar,Oeschler}, the role of the
$4f$-electron hybridization with the conduction electrons in the
occurrence of superconductivity\cite{AndyCEF,NeilHarrison,Raj} and
the effects of quasi-2D electronics
structures\cite{Hall,Hall2,quintana}are some of the physical
phenomena that have been brought to the scenario to answer the
question above. Further, motivated by this experimental trend, new
materials search based on the 1-1-5 structures has led to the
discovery of the Pu-based HFS PuMGa$_{5}$ (M = Rh and
Co).\cite{Sarrao,eric}

On the other hand, as these HFS are presumably
magnetically mediated, others
studies\cite{Kumar,Andy2,pagliuso3,pagliuso4,raimundo1,raimundo2,japa115,Malinowski,Victor,granado1,pagliuso5}
have been focused in understanding the evolution of the $4f$ local
magnetism, not only for the magnetically ordered Ce-based members
of this family such as CeRhIn$_{5}$ and Ce$_{2}$RhIn$_{8}$, but
also for their antiferromagnetic counterparts R$_{m}$M$_n$In$_{3m+2n}$
(M = Rh or Ir, $m$ = 1, 2;) for $R$ =  Nd, Sm, Gd and Tb. From
these studies, it was established the role of tetragonal crystalline
electrical field (CEF) in determining the spatial direction of the ordered
R-moments with respect to the lattice and the evolution of the
N\'{e}el temperature, $T_N$, in the
series.\cite{pagliuso3,pagliuso4,raimundo1,raimundo2,pagliuso5}

A key set of experiments allowing the above conclusions was the experimental determination of the magnetic structures of
various members of the R$_{m}$M$_n$In$_{3m+2n}$ (M = Rh or Ir, $m$ =
1, 2;)
family.\cite{wei,wei1,wei2,Andy2,raimundo2,granado1,granado2} Up to date, however, none of the Sm-based compounds from this family have had their magnetic structures determined. In
fact, the compounds of this series containing Sm ions may be particularly important
in testing the extension of the CEF trends in this family because the presence of excited
$J$-multiplet states in Sm$^{3+}$ and quadrupolar interactions have to be taken
into account in order to understand their magnetic phase
diagrams.\cite{kasaya,endoh,Kletowski,Stunault} Especially interesting
is Sm$_2$IrIn$_8$ which presents a first order antiferromagnetic
transition at $T_N$ = 14.2 K.\cite{pagliuso3} This value is
slightly smaller than the $T_N$ $\sim$ 16 K of the cubic
SmIn$_{3}$\cite{buschow} which according to the CEF trends
observed in other members of this
family\cite{raimundo2,pagliuso5} suggest that the ordered
Sm-moments should lie the $ab$-plane.

To further explore the magnetic properties of Sm$_2$IrIn$_8$ and to check the extension of the CEF trends
observed for R = Nd, Tb, and Ce,\cite{pagliuso3,pagliuso4,raimundo1,raimundo2,pagliuso5} to the Sm-based compounds, we report in this work the solution of the magnetic
structure of the intermetallic antiferromagnet Sm$_2$IrIn$_8$ by means of the x-ray resonant magnetic scattering
(XRMS) technique. The XRMS technique has proved to be a very important tool for the investigation of microscopic magnetism in
condensed matter, specially for highly neutrons absorber ions such as Sm.

Sm$_2$IrIn$_8$ presents, below $T_N$ = 14.2 K, a commensurate
antiferromagnetic structure with a propagation vector $\vec{\eta}
= (\frac{1}{2},0,0)$. The Sm magnetic moments lie in the
\textit{ab} plane. In terms of relative orientation, the propagation vector $\vec{\eta}$ indicates that the Sm-spins are
ordered antiferromagnetically along the \textit{a} axis and
ferromagnetically along the \textit{b} axis and, because of the presence of two Sm ions per unit cell along \textit{c} axis, some calculations have to be performed in order to determine the type of ordering along this direction. Furthermore, as
it could be expected for such spin arrangement in a tetragonal
compound, antiferromagnetic domains were observed in the
ordered state of Sm$_2$IrIn$_8$. These domains were removed by
field-cooling the sample at a field of $H$ = 10 T.

\section{\bf EXPERIMENT}

Single crystalline samples of Sm$_2$IrIn$_8$ were grown from Indium flux as described
previously.\cite{Fisk2,pagliuso3} The crystal structure, unit cell dimensions and macroscopic properties of the
Sm$_2$IrIn$_8$ single crystals used in this work were in agreement with the data in Ref. \onlinecite{pagliuso3}.
For the XMRS experiments of this work, selected crystals were extracted and prepared with polished (0,0,$l$)
flat surfaces, and sizes of approximately 4 mm x 3.4 mm x 1.5
mm. The preferred crystal growth direction of this tetragonal compound is columnar along the [00\textit{l}] direction and the (001)
facet is relatively large. The mosaic spread of the sample was found to be $<0.08$° by a rocking curve ($\theta$ scan) on a Phillips four circle diffractometer.

XRMS studies were performed at the 4-ID-D beamline at the Advanced
Photon Source (APS) and at the ID-20 beamline at the European
Synchrotron Radiation Facility (ESRF). The 4-ID-D x-ray source is a 33 mm period planar undulator and the energy is selected with a double crystal
Si(111) monochromator. A toroidal mirror focuses the beam to a 220 $\mu$m (horizontal) x 110 $\mu$m (vertical) spot, yielding an incident flux of $\sim$3.5 x 10$^{13}$ photons/s with an energy resolution of $\delta E/E$ = 1.4 x 10$^{-4}$. The sample was cooled in a
closed-cycle He refrigerator (with a base temperature of 4 K) with
a dome Be window. Our experiments were performed in the coplanar
geometry with $\sigma$-polarized incident photons, i.e., in the
vertical scattering plane, using a four-circle diffractometer. Except for azimuthal scans, the sample was mounted with the \textit{b} axis perpendicular to the scattering plane.

In most measurements, we have performed a polarization analysis,
whith Cu(220), Graphite (006) and Au(111) crystal analysers, appropriate for the energies of Sm $L_2$ and
$L_3$ edges. The diffractometer configuration at the APS allowed measurements at different azimuthal angles ($\phi$) by rotating the sample around the scattering
vector \textbf{Q}. This was particularly useful to probe the magnetic moment components at the dipolar resonant condition with $\sigma$ incident polarization. 

The x-ray source on the ID-20 beamline was a linear undulator
with a 32 mm period. The main optical components are a double Si(111) crystal monochromator with sagital focusing 
and two meridional focusing mirrors on either side of the monochromator. At
7.13 keV using the first harmonic of the undulator u32, the standard incident flux at the sample position was
approximately 1 x 10$^{13}$ ph/s at 200 mA with a beam size of  500 $\mu$m (horizontal) x 400 $\mu$m (vertical). The sample was mounted on a
cryomagnet (with a base temperature of 2 K), installed on a
horizontal six-circle diffractometer, with the \textit{b} axis
parallel to the cryomagnet axis and perpendicular to the
scattering plane. This configuration allowed $\pi$-polarized incident
photons in the sample and the application of an external magnetic
field up to 10 T perpendicular to the scattering plane.

\section{\bf RESULTS}

\subsection{Temperature dependence and resonance analysis}

\begin{figure}
\centering
        \includegraphics[width=0.48 \textwidth]{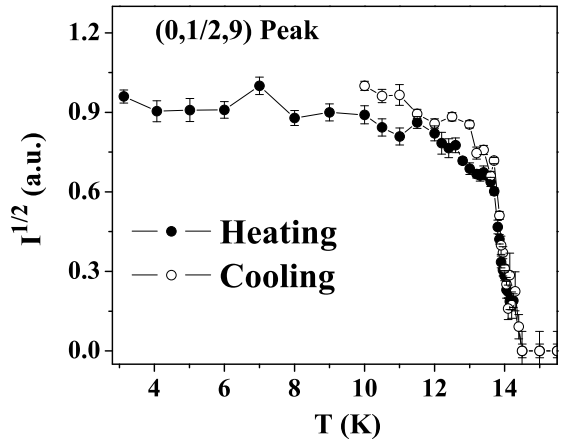}
\caption{Temperature dependence of one Sm$_2$IrIn$_8$ sub-lattice magnetization measured with transverse ($\theta$) scans at the ($0,\frac{1}{2},9$) peak.}\label{fig:TempDepend}
\end{figure}

Magnetic peaks were observed in the dipolar resonant condition at temperatures below $T_N$ = 14.2 K at reciprocal lattice points forbidden for charge scattering and consistent with an antiferromagnetic structure with propagation vector $(\frac{1}{2},0,0)$. Their temperature dependence was studied for increasing  and decreasing temperature sweeps. Figure~\ref{fig:TempDepend} shows the temperature dependence of
($0,\frac{1}{2},9$) magnetic reflection at an incident photon energy of 7.313 keV ($L_2$ edge) and measured at $\pi$ incident polarization without polarization analysis. The squared root of the integrated intensity, which is proportional to a Sm sub-lattice magnetization, is displayed. A
pseudo-voigt peak shape was used to fit transversal $\theta$ scans
through the reciprocal lattice points in order to obtain the integrated
intensities of the reflection peak. This peak intensity decreases
abruptly to zero for T $>$ 13 K and its critical behavior can not
be described by a power-law function with a critical exponent $\beta$. This
result is in agrement with the first order character of the
magnetic transition at 14.2 K, revealed by heat capacity data,
from which a latent heat of $\sim{10}$ J/mol was
extracted.\cite{pagliuso3} Consistently, we found evidence of
small hysteresis for T $\lesssim$ 14.2 when changing from warming
to the cooling temperature sweep.

\begin{figure}
\centering
        \includegraphics[width=0.48 \textwidth]{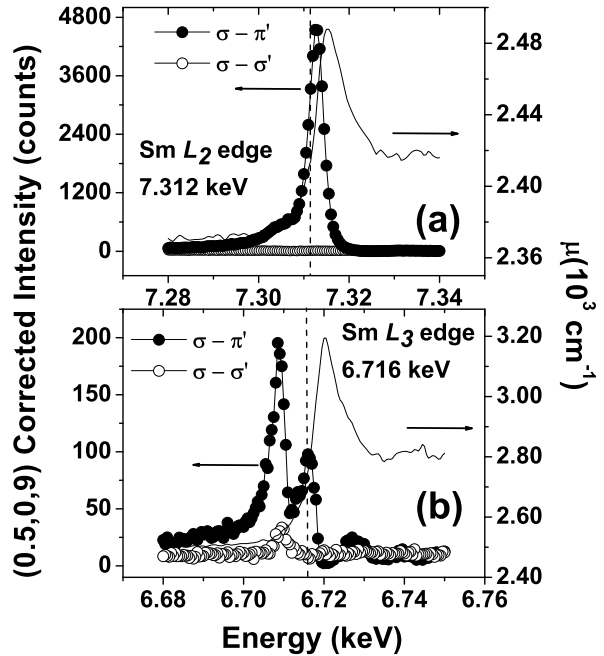}
\caption{Energy scan of the ($\frac{1}{2}$,0,9) magnetic peak at T = 5.9 K for $\sigma$ - $\pi$' (closed circles) and $\sigma$ - $\sigma$' (open circles) polarization channels at the $L_2$ (top) and $L_3$ (bottom) absorption edges. The data have been corrected for absorption, $\mu (E)$,
using the measured fluorescence yield. Arrows indicate the scales for the fluorescence yield (right) and the observed data (left).}\label{fig:EnergyScans}
\end{figure}

The energy line shape curves for the polarization channels $\sigma$ - $\pi$' and $\sigma$ - $\sigma$'  of the ($\frac{1}{2}$,0,9) diffraction peak at (a) the
$L_2$ and (b) the $L_3$ absorption edges of Sm$^{3+}$ ion at $T$ = 5.9 K are shown in
Figure~\ref{fig:EnergyScans}. The solid lines in both panels represent the absorption spectrum, $\mu (E)$,
extracted from fluorescence yield. The data of Figure~\ref{fig:EnergyScans} were collected at the 4-ID-D beamline
of APS by counting the photons reaching the detector at a fixed \textbf{Q} while changing the incident energy. The strong resonant enhancement of the x-ray scattering at this reciprocal space position provide clear evidence of the magnetic origin of the observed peaks.

The energy scan curve in Figure~\ref{fig:EnergyScans}(a) has a maximum at
7.312 keV which is only $\sim$2.5 eV larger than the $L_2$
absorption edge (defined by the inflection point of the absorption
spectrum), revealing the electric dipolar character (\textit{E}1)
of this transition (from 2\textit{p} to 5\textit{d} states).
Figure~\ref{fig:EnergyScans} also shows the polarization analysis
 performed to unambiguously confirm the magnetic
origin of the superstructure peaks. Polarization
analysis was also used to verify whether the anomaly at approximately 8
eV below the dipolar peak in Figure~\ref{fig:EnergyScans}(a) could
be associated with a quadrupolar transition\cite{Hill} or it
simply represents an enhanced interference between the
non-resonant and the resonant part of the scattering amplitude.
For the experimental configuration used (incident
$\sigma$-polarization), the electric dipole transitions
\textit{E}1 rotate the plane of polarization into the scattering
plane ($\pi$-polarization). Our data in
Figure~\ref{fig:EnergyScans}(a) reveals a strong enhancement of the
scattered intensities at the $\sigma$ - $\pi$' channel (closed
circles) and no enhancement at the $\sigma-\sigma$'
channel for the same energy range. These results confirm the magnetic origin of the
$(h,0,l$)$ \pm (\frac{1}{2},0,0)$ reflections due to the existence
of an antiferromagnetic structure doubled along the
crystallographic $\hat{a}$ direction, with a propagation vector
$\vec{\eta} = (\frac{1}{2},0,0)$.

The energy profile around the Sm
$L_3$ edge is presented in Figure~\ref{fig:EnergyScans}(b). Firstly, the observed
intensities are roughly one order of magnitude weaker than those
obtained at the $L_2$ resonance, in agreement with previous measurements on pure Sm.\cite{Stunault} Secondly, there are two peaks in
the $\sigma$ - $\pi$' channel signal, as also observed for other
light rare-earth\cite{Zheludev,Hill1} and Sm-based
compounds.\cite{Stunault,Detlefs} A high energy peak appears at
6.716 keV, while a low energy and more intense enhancement can be
observed at 6.708 keV. Interestingly, Stunault \textit{et
al.}\cite{Stunault} have demonstrated that for pure Sm the quadrupolar
\textit{E}2 resonance is more intense than the dipolar \textit{E}1
 at the $L_3$ edge and they found that the energy
difference between the \textit{E}2 and the \textit{E}1 resonances is of the order of 8 eV, the same as the one found in this work. Furthermore, in the $\sigma$ - $\sigma$' channel only an enhancement at 6.708 keV could be observed which is consistent with the
quadrupolar character of this resonance, since scattering signal
in $\sigma$ - $\sigma$' channel for dipolar transitions is strictly
forbidden.\cite{Hannon,Hill} Thus, the presence
of this pre-edge enhancement in the energy curves of
Figure~\ref{fig:EnergyScans} confirms an expected quadrupole (\textit{E}2) 
2\textit{p} to 4\textit{f}  contribution to the
resonant x-ray scattering in Sm$_2$IrIn$_8$.

\subsection{The magnetic structure}

The magnetic structure of the Sm$_2$IrIn$_8$ was experimentally investigated using dipolar resonant x-ray magnetic scattering with polarization analisys. In general, the magnetic scattering intensities are given by:\cite{Detlefs,Hill} 

\begin{eqnarray} \label{eq:equation1}
I\propto\frac{1}{\mu^{*}sin(2\theta)}\left|\sum_{n}\textit{f}_{n}^{XRES}(\vec{k},\hat{\epsilon},\vec{k'},\hat{\epsilon'})e^{i\vec{Q} \cdot \vec{R}_n}\right|^{2},
\end{eqnarray} 
where $\mu^{*}$ is the absorption correction for asymmetric reflections, 2$\theta$ is the scattering angle, $\vec{Q}=\vec{k'}-\vec{k}$ is the wave-vector transfer, $\vec{k}$ and $\vec{k'}$ ($\hat{\epsilon}$ and $\hat{\epsilon'}$) are the incident and scattered wave (polarization) vectors, respectively. $\vec{R}_{n}$ is the position of the \textit{n}th resonant atom in the lattice, and $\hat{z}_{n}$ is the moment direction of this atom. The resonant scattering amplitude  contains both dipole (\textit{E}1) and quadrupole (\textit{E}2) contributions. For the determination of the magnetic structure of this work we have used the second term of the electric dipole transition (\textit{E}1) form factor which produces magnetic peaks. In this case we have:

\begin{widetext}
\begin{eqnarray} \label{eq:equation2}
\textit{f}_{nE1}^{XRES} \propto \left[ \begin{array}{cc}
0 & \hat{k}'\cdot\hat{z}_{n} \\
-\hat{k'}\cdot\hat{z}_{n} & (\hat{k}\times\hat{k'})\cdot\hat{z}_{n}\\
\end{array} \right] \propto \left[ \begin{array}{cc}
0 & z_1 cos \theta + z_3 sin \theta\\
- z_1 cos \theta + z_3 sin \theta & -z_2 sin (2\theta)\\
\end{array} \right],
\end{eqnarray}
\end{widetext}

where $\theta$ is the Bragg angle, $z_1$, $z_2$ and
$z_3$ are the components of the magnetic moment at the
\textit{n}th site, according to the commonly used geometry
convention of Ref. \onlinecite{Gibbs1}; $\sigma$, $\pi$, $\sigma$'
and $\pi$' describe the incident (non-primed terms) and scattered
(primed) photon polarizations.

As described previously, two experimental setups have been used in
this work, in the vertical (4-ID-D beamline) and horizontal
(ID-20) scattering configurations. This permitted us to access all four polarization channels of the 2x2 matrix
in~(\ref{eq:equation2}) and to determine the magnetic moment orientations through their polarization dependence at the
\textit{E}1 resonance by comparing the relative intensities of
experimental $(\frac{1}{2},0,l)$ magnetic peaks with the
calculated ones using the appropriate terms of
matrix~(\ref{eq:equation2}).\cite{Detlefs}

\begingroup
\begin{table*}
\begin{center}
\caption{\label{tab:table1}Comparison between observed and
calculated intensities of magnetic Bragg reflections, assuming
either parallel (model I) or antiparallel (model II) alignment
between the moments of two Sm ions along the \textit{c} axis in the same
chemical unit cell.}
\begin{ruledtabular}
\begin{tabular}{cccccc}
 &\multicolumn{3}{c}{\hspace{3cm} MODEL I}&\multicolumn{2}{c}{MODEL II}\\
($h,k,l$) & Exp. Data & \textbf{m}//\textit{c}& \textbf{m}//\textit{a} & \textbf{m}//\textit{c}& \textbf{m}//\textit{a} \\
 (1/2,0,6) & 66 & 13 & 29 & 24 & 55 \\
 (1/2,0,7) & 78 & 17 & 29 & 39 & 68 \\
 (1/2,0,8) & 5 & 77 & 100 & 3.4 & 4.5 \\
 (1/2,0,9) & 100 & 3 & 3 & 100 & 100 \\
 (1/2,0,10) & 12 & 100 & 68 & 32 & 23 \\
\end{tabular}
\end{ruledtabular}
\end{center}
\end{table*}
\endgroup

\begin{figure}
\centering
        \includegraphics[width=0.42 \textwidth]{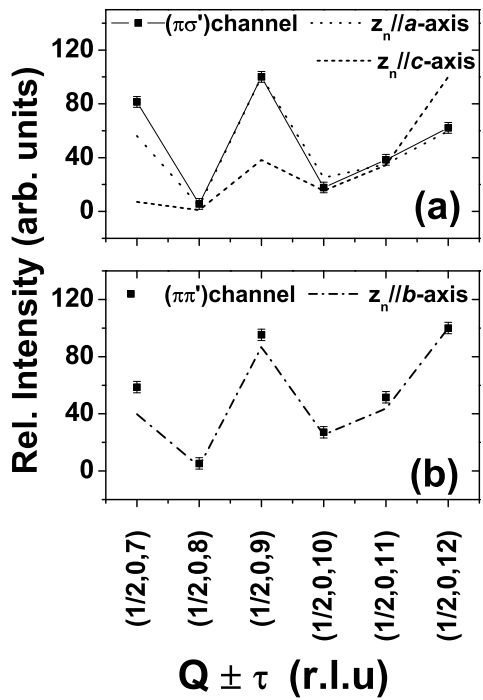}
\caption{Analysis of the possible magnetic moment directions
 for Sm$_2$IrIn$_8$ at the $L_2$ resonance.
\textbf{Q}-dependence of the integrated intensities of: (a) six satellite peaks signal in the $\pi - \sigma'$ channel with the moments along $\hat{a}$ and $\hat{c}$, and (b) in the $\pi - \pi'$ with moments parallel to
$\hat{b}$.}\label{fig:IntensInteg}
\end{figure}

In the case of Sm$_2$IrIn$_8$ the magnetic propagation vector $\vec{\eta} = (\frac{1}{2},0,0)$ does not
unequivocally determine the magnetic structure due to the presence of two magnetic Sm atoms per chemical unit
cell along the $\hat{c}$ direction. Therefore, as stated above, we have an antiparallel ordering of the Sm moments along the $\hat{a}$ direction and a parallel ordering along $\hat{b}$. Along $\hat{c}$ there are, however, two possibilities of coupling that can take place: a parallel arrangement (Model I), in which the moments of neighboring Sm ions along \textit{c} axis are parallel to
each other (sequence $+\!+\:+\!+\:+\!+\: \dots$), or the antiparallel coupling (Model II), with the sequence
($+\!-\:+\!-\:+\!-\:\dots$). These two possibilities have been considered into the calculated magnetic structure
factor while orienting the magnetic moment along the three crystallographic directions for five different $(\frac{1}{2},0,l)$ magnetic Bragg peaks, with \textit{l} = 6, 7, 8, 9, 10. The calculated intensities are strongly dependent on the projections
of magnetic moments along the crystallographic axis through the product $\hat{k}'\cdot\hat{z}_{n}$ of equation~(\ref{eq:equation2}). Therefore, they were compared to the relative observed intensities for each case. This
evaluation was performed at the vertical geometry of the 4-ID-D beamline at 9 K by performing rocking scans
with the crystal analyzer and numerically integrating the data.\cite{Detlefs} We show this analysis in
Table~\ref{tab:table1}, where ``Model I'' stands for the $+\!+\:+\!+\:+\!+\:\dots$ sequence and ``Model II'' for
the $+\!-\:+\!-\:+\!-\: \dots$ one. This comparison shows that the model which best fits the experimental data is the one assuming antiparallel coupling along \textit{c} axis (Model II) with the magnetic moments approximately oriented along the \textit{a} axis (according to
matrix~(\ref{eq:equation2}), for a $\sigma$ polarized incident beam and peaks at reciprocal space positions with the
(001) normal surface contained in the scattering plane, contributions from an oriented moment along $\hat{b}$
direction cannot be detected). 

In addition, we have also measured the $\pi - \sigma'$ and $\pi - \pi'$ polarization channels at the horizontal
geometry of the ID-20 beamline. Measuring these two channels we gained access to the $z_1$ and $z_3$ components (in equation 2)
of magnetic moment vector in one case [$\pi - \sigma'$, Figure~\ref{fig:IntensInteg}(a)] and to $z_2$ in the other
[$\pi - \pi'$, Figure~\ref{fig:IntensInteg}(b)]. There is a clear indication that for the $\pi - \sigma'$ channel
the observed data are well fit when considering the moments along the $\hat{a}$ direction [dotted curve in
Figure~\ref{fig:IntensInteg}(a)] instead of $\hat{c}$ direction [short dashed curve]. Also in this case the
\textit{E}1 terms are not sensitive to the component of the ordered moment perpendicular to the scattering
plane, i.e. along \textit{b} axis. Further, when measuring the channel ($\pi - \pi'$) we are only
allowed to measure the \textit{b} component, which is confirmed by the good fit of experimental data when
assuming magnetic moments along such direction [dash-dotted curve in Figure~\ref{fig:IntensInteg}(b)]. These two
last results indicate that the Sm moments actually have components along both \textit{a} and \textit{b} real
space axis and not perfectly aligned along any of these two directions.

\begin{figure}
\centering
        \includegraphics[width=0.48 \textwidth]{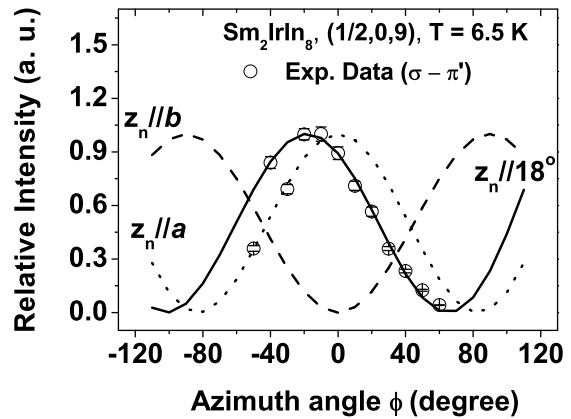}
\caption{Azimuth scan analysis. Normalized integrated intensities of the ($\frac{1}{2},0,9$)
 magnetic peak at T = 6.5 K (open circles). The other curves
represent the integrated intensities behavior considering the magnetic moments along
the $\hat{a}$ (dotted line), $\hat{b}$ (dashed) and 18º away from
$\hat{a}$ (solid line) direction.}\label{fig:Azimuth}
\end{figure}

\begin{figure}
\centering
        \includegraphics[width=0.45 \textwidth]{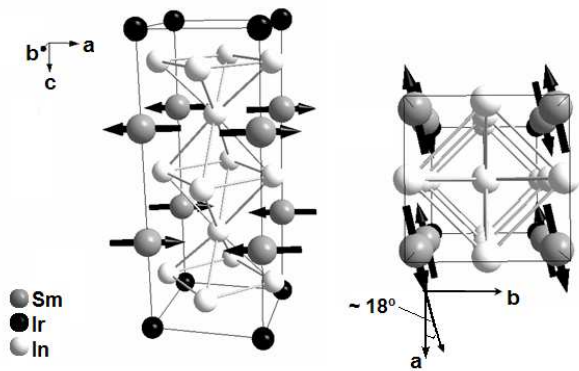}
\caption{Magnetic structure of Sm$_2$IrIn$_8$ below $T_N$ = 14.2 K (left) and a Sm-In plane top view (right) showing the in-plane arrangement of Sm moments.}\label{fig:MagStruct}
\end{figure}

To determine the exact orientation of the magnetic moments within
the \textit{ab} plane, we have performed azimuthal scans ($\phi$
scan) through the ($\frac{1}{2},0,9$) reflection
(Figure~\ref{fig:Azimuth}) at the \textit{E}1 resonance. At the
$\sigma - \pi'$ polarization channel this procedure warrants the
determination of moments directions with no ambiguity because the
magnetic cross section is strongly dependent of the magnetic
moment direction and the polarization of the incoming and
scattered radiation, the maximum (minimum) intensity in the curve
will occur with the magnetic moment being parallel
(perpendicular) to the diffraction plane. With the experimental
setup of 4-ID-D beamline we had access to record points at
azimuthal angles $\phi$ between -50º and 60º. In order to compare
with the observed data, one can calculate the intensities for the
$\sigma - \pi'$ channel using the expressions~(\ref{eq:equation1})
and~(\ref{eq:equation2}) and a reasonably simple geometry analysis
considering the projections of both $\hat{k'}$ and $\hat{z_n}$ on
the coordinate system of Ref. \onlinecite{Gibbs1} when the azimuth
angle is changed. Then, the calculated intensity is proportional to
$I^{\sigma\pi'}\propto |$-cos$\theta$ cos$\phi$ cos$\alpha$ +
sin$\theta$ sin$\alpha |^2$, where $\alpha$ represents the
assymetry angle between the scattering and the normal surface
vector.\cite{Detlefs} Figure~\ref{fig:Azimuth} shows the
experimental and the calculated relative intensities
considering the moment along the \textit{a} and \textit{b} axis, as well as
18º tilted from the \textit{a} axis, which is the value that
nicely adjust the experimental data. Considering the experimental
errors we can then conclude that the magnetic moment is in the
\textit{ab} plane making (18º $\pm$ 3º) with the $\hat{a}$
direction of the sample. Using all the above results, a model of the
magnetic unit cell of Sm$_2$IrIn$_8$ can be constructed and is shown in
Figure~\ref{fig:MagStruct}.

As it was observed in the magnetic structure of other members of
the R$_{m}$MIn$_{3m+2}$ series such as NdRhIn$_{5}$\cite{wei}, TbRhIn$_5$,\cite{raimundo2} GdRhIn$_5$,\cite{granado2} and
Gd$_{2}$IrIn$_{8}$\cite{granado1} the magnetic structure of
Sm$_2$IrIn$_8$ presents a lower symmetry than the crystallographic
structure, as the Sm spins present different relative orientations along the
$\hat{a}$ and $\hat{b}$ directions even though $a$ and $b$ are
indistinguishable. This spin arrangement was explained by
considering the first ($J_1$) and second ($J_2$) R-neighbors
exchange interactions in the case of a small $J_1$/$J_2$
ratio.\cite{granado2}

Considering the observation of this kind of magnetic structure in
tetragonal compounds, it may be expected that at zero magnetic
field the antiferromagnetic ordering takes place with the
formation of antiferromagnetic domains where the relative
orientation of the magnetic moments along a given direction ($\hat{a}$
or $\hat{b}$) changes from parallel to antiparallel between the domains. The presence of a twinned magnetic structure with symmetry-related domains was evidenced by the observation of both ($\frac{1}{2},0,l$) and ($0,\frac{1}{2},l$) reflection-types in this work. To further investigate the presence of antiferromagnetic domains in the
ordering state of Sm$_2$IrIn$_8$ we follow the behavior of the magnetic ($\frac{1}{2},0,l$) and ($0,\frac{1}{2},l$) reflections
under an applied magnetic field.

\begin{figure}
\centering
        \includegraphics[width=0.48 \textwidth]{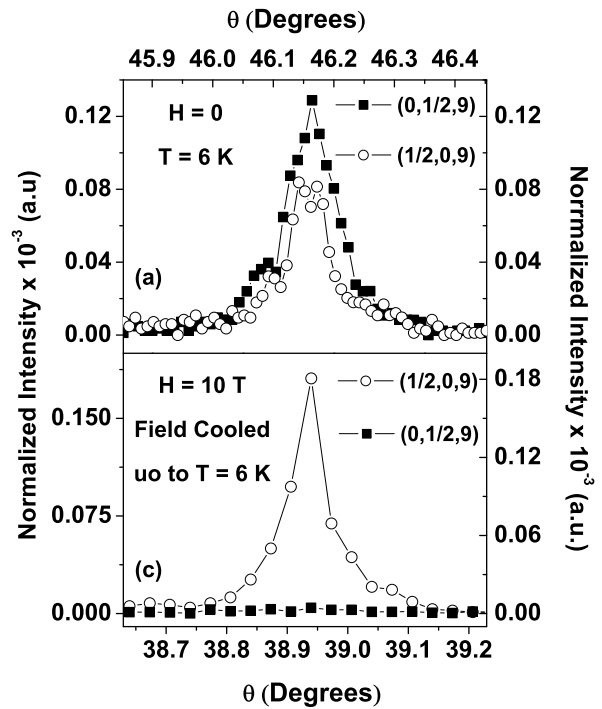}
\caption{Field-dependence of the integrated intensities of the ($\frac{1}{2},0,9$)
 and ($0,\frac{1}{2},9$) magnetic peaks taken with transverse
($\theta$) scans around each reciprocal space lattice points. (a)
For H = 0 applied field at T = 6 K, (b) for H = 10 T and (c) field cooled from 16 K to 6 K at H = 10
T.}\label{fig:AppliedField}
\end{figure}

Figure~\ref{fig:AppliedField} presents the behavior of the ($\frac{1}{2},0,9$) and ($0,\frac{1}{2},9$) intensities as a function of the applied magnetic field of 10 T along one of the tetragonal axis in the plane (defined as $\hat{b}$ direction). At zero field and $T$ = 6 K, both ($\frac{1}{2},0,9$) [open circles] and ($0,\frac{1}{2},9$) [closed squares] intensities can be observed with comparable magnitude [Figure~\ref{fig:AppliedField}(a)]. The ($\frac{1}{2},0,9$) intensity is roughly 66\% that of the ($0,\frac{1}{2},9$) peak. The sample was then field cooled ($H$ = 10 T) from the paramagnetic (16 K) to the ordered state (6 K) with the field applied along the $\hat{b}$ direction. As can be seen in Figure~\ref{fig:AppliedField}(b) the ($0,\frac{1}{2},9$) diffraction peak disappears as the magnetic field favors the parallel spin orientation along the $b$ axis. The same effect was also observed for the other five ($0,\frac{1}{2},l$) reflections (not shown). The results under applied magnetic field shown in
Figure~\ref{fig:AppliedField} confirm the existence of a twinned magnetic structure for Sm$_2$IrIn$_8$ which
allows the observation of both ($0,\frac{1}{2},l$) and ($\frac{1}{2},0,l$) magnetic reflections at zero field.

\section{\bf DISCUSSION}

Early studies on the antiferromagnetic cubic compound
SmIn$_3$ have shown multiple magnetic transitions associated with
quadrupolar ordering, magnetoelastic and magnetocrystalline
competitive effects at 14.7, 15.2 and 15.9 K (the former two
temperatures being associated with successive magnetic dipolar,
antiferromagnetic, orders and the last one due to quadrupolar
ordering).\cite{kasaya,endoh} For the tetragonal Sm$_2$IrIn$_8$,
the insertion of two additional SmIn$_3$ atomic layers into the crystalline
structure slightly decreases $T_N$ compared to that of SmIn$_3$
(14.2 and 15.2 K for the Sm2-1-8 and Sm1-0-3 $T_{N}$'s,  respectively) and an additional anomaly at
11.5 K has been observed in the specific heat and resistivity
measurements,\cite{pagliuso3} probably related to the successive
transitions seen in the ordered phase of the SmIn$_3$.

Following the investigation of the isostructural magnetic non-Kondo
compounds from the R$_{m}$MIn$_{3m+2}$ family, where the details
the 4$f$ magnetism along the series may be important to understand
the possible magnetic-mediated superconductivity in the compounds
with \textit{R} = Ce, we have studied the
magnetic structure of Sm$_2$IrIn$_8$, which is the only compound
from this family with a clear first order antiferromagnetic
transition and now it is the first Sm-member from this family with a solved
magnetic structure, which is the main result of this work. The determination of the Sm2-1-8 magnetic structure allows for the investigation of the CEF driven trends of magnetic properties within the R$_{m}$MIn$_{3m+2}$ family to be extended to the Sm-based members. 

Our results confirm the complex resonance profile of Sm-based
compounds (at one satellite reciprocal point, Figure~\ref{fig:EnergyScans}), as seen in previous studies of pure Sm.\cite{Stunault} It has been argued that the larger intensity of \textit{E}2 resonance at Sm $L_3$ edge compared
to its intensity at the $L_2$ edge may be explained qualitatively by the spin-orbit splitting of the
intermediate 4\textit{f} levels involved.\cite{Stunault} The $L_3$ transitions connect the $j = \frac{7}{2}$
state while $L_2$ involves transitions to the $j = \frac{5}{2}$ level, which lie lower in energy and therefore
can be preferentially populated by the five 4\textit{f} Sm electrons. This reduces the number of vacant $j =
\frac{5}{2}$ states from 6 to 1, in contrast to the 8 states available for the $j = \frac{7}{2}$ level, which
increases the transition probability of the \textit{E}2 resonance at Sm $L_3$ in Sm$_2$IrIn$_8$.

Considering the additional magnetic transitions observed for SmIn$_3$,\cite{kasaya,endoh} and the additional anomaly at $T$ = 11.5 K in heat capacity and electrical resistivity measurements for Sm$_2$IrIn$_8$,\cite{pagliuso5} we did not observe any discontinuities, within the resolution of our experiment, in the integrated intensities of the ($0,\frac{1}{2},9$) magnetic peak from roughly 4 K up to 16 K (Figure~\ref{fig:TempDepend}). Therefore we conclude that there are no changes of the magnetic propagation vector $\vec{\eta} = (\frac{1}{2},0,0)$ below $T_N$. For completeness, on going field-dependent heat capacity and thermal expansion measurements (not shown and will be published elsewhere) have revealed no field-induced transitions up to \textit{H}=9 and 18 T, respectively, similarly to SmIn$_3$ where no additional transition was found with applied field up to \textit{H}=32 T.\cite{Kletowski}

On the other hand, recent works have shown that the low temperature CEF configuration plays
a fundamental role on the behavior of $T_N$ and the magnetic
moment directions within the R$_{m}$MIn$_{3m+2}$ family.\cite{wei2,pagliuso5,raimundo1,raimundo2} Further, Kubo \textit{et al.}\cite{JapaCEF} has also  proposed an orbital controlled mechanism for superconductivity in the Ce-based compounds from this family. For the Sm
members, CEF effects confine the magnetic moments to the \textit{ab} plane, consistent with the experimental CEF trends observed for R = Ce, Nd and Tb\cite{pagliuso3,pagliuso4,raimundo1,raimundo2} and also by
the predictions of a recently developed mean field theoretical
model.\cite{pagliuso5,raimundo2} If the magnetic ordered moments lie in the
$ab$-plane but they are more magnetically susceptible along the
$c$ axis the magnetic order can be frustrated to lower $T_N$
values than for their cubic relatives. The mean-field model of Ref. \onlinecite{pagliuso5}, however, only includes the contributions of
tetragonal CEF and first neighbor isotropic dipolar exchange
interaction. Therefore, it may not be expected to work for
Sm containing compounds, because for the Sm$^{3+}$ ion the first excited
\textit{J}-multiplet lying just above the ground state is closer
in energy. Thus, the tetragonal CEF splitting can mix both the
excited and ground state CEF scheme and this particular effect
should be considered into the calculations. Indeed, this is the
responsible for the non-linear response of the inverse of magnetic
susceptibility at high temperatures on SmIn$_3$ and other Sm-based
compounds,\cite{buschow,Tsuchida} as well as in Sm$_2$IrIn$_8$.\cite{pagliuso3}
Furthermore, as it was found for SmIn$_3$,\cite{kasaya,endoh}
quadrupolar magnetic interactions also have to be considered in
order to achieve a complete description of the magnetic properties
of the Sm-based compounds in the R$_{m}$MIn$_{3m+2}$ family.

Apart from the higher complexity of the magnetic properties
of the Sm-compounds, it was found experimentally that $T_N$ is
decreased (roughly $\sim{10}\%$) for the tetragonal compounds when
compared to the cubic SmIn$_3$. In addition, we have found that the magnetic structure of Sm$_2$IrIn$_8$ shows the ordered Sm moments in the $ab$ plane, as expected in the case of $T_N$ suppression.\cite{pagliuso5,raimundo2} Although the changes in $T_N$ for the Sm compounds are much smaller
(perhaps due to the particularities of the Sm$^{3+}$ ion discussed above) than that observed for R = Ce, Nd and Tb in the R$_{m}$MIn$_{3m+2}$ family, we can conclude with the solution of the magnetic structure reported here, that the general CEF trend of the R$_{m}$MIn$_{3m+2}$ is also qualitatively present in Sm$_2$IrIn$_8$.

\section{\bf CONCLUSION}

In summary, we have presented the results of the magnetic
structure determination of the intermetallic antiferromagnet
Sm$_2$IrIn$_8$. The magnetic order is commensurate with
propagation vector $\vec{\eta} = (\frac{1}{2},0,0)$ and the Sm
moments oriented in the \textit{ab} plane. We used different
scattering geometries (exploring the polarization dependences of
magnetic intensities) and azimuthal scans around a magnetic
reciprocal space point to determine without ambiguity that the
moments are aligned approximately 18º away from the \textit{a}
axis. The temperature behavior of the magnetic satellites have
been probed at the ($0,\frac{1}{2},9$) reciprocal node and show no
evidence of changes in the magnetic structure within the studied
temperature range. Besides,  an abrupt (non-power law) decrease of
magnetic intensities at  $T_N$ was found, consistent with the
first order character of the antiferromagnetic transition of
Sm$_2$IrIn$_8$. The resonance properties at the Samarium $L_2$ and
$L_3$ absorption edges revealed both resonant \textit{E}1 and
\textit{E}2 process with roughly one order of magnitude more
intense resonance peaks at the $L_2$ edge and a much stronger
quadrupole resonance in the $L_3$ edge. The orientation of Sm
moments in the \textit{ab} plane and the small decrease of $T_N$
compared to its value for SmIn$_3$ agrees with a general CEF trend found in
the R$_{m}$MIn$_{3m+2}$ family.

\begin{acknowledgments}
This work was supported by FAPESP (SP-Brazil) Grants No.
05/55272-9, 05/00962-0, 04/08798-2 and 03/09861-7, CNPq (Brazil)
Grants No. 307668/03, 04/08798-2, 304466/20003-4 and
140613/2002-1, and FAEPEX (SP-Brazil) Grant No. 633/05. Use of the Advanced
Photon Source was supported by the U. S. Department of Energy, Office of Science, Office of Basic Energy
Sciences, under Contract No. DE-AC02-06CH11357. The staff at the 4-ID-D and ID-20 beam lines are gratefully acknowledged for
providing an outstanding scientific environment during these experiments.

\end{acknowledgments}

\bibliography{Bibliog}

\end{document}